# Title of PhD Project: Improving the safety and quality of medical imaging by bridging the gap between health information technology and human performance

PhD Student
MD Shafiqur Rahman Jabin
The University of South Australia
rahmy027@mymail.unisa.edu.au

Principal Supervisor
William B. Runciman
The University of South Australia
William.Runciman@unisa.edu.au

Associate Supervisor
Tim Schultz
The University of Adelaide
tim.schultz@adelaide.edu.au

Associate Supervisor
Peter Hibbert
The University of South Australia
peter.hibbert@unisa.edu.au

Associate Supervisor
Catherine Mandel
The University of Melbourne
mandel@ausdoctors.net

## MOTIVATION AND RESEARCH QUESTIONS

The utility and scope of medical imaging has expanded over the last 20 years and is now a critical part of medical practice. At the same time, Health Information Technology (HIT) has emerged as an indispensable tool for medical imaging that has been deployed widely. The benefits of HIT include collaborative and differential diagnosis; weighing and accuracy of diagnosis; and facilitation of diagnostic feedback and workflows. Although HIT has the potential to provide advantages for healthcare delivery and patient outcomes, it can pose substantial risks to patient safety if poorly designed, implemented or managed. Such risks may not only be harmful, but difficult to detect. Healthcare organizations in Australia are experiencing the same types of incidents reported elsewhere and being increasingly recognized.

Our research questions are categorized in sections. Firstly, what are the main patient safety risks in radiology? Secondly, what are the proposed solutions to these? What interventions, particularly those related to HIT may reduce incidents in the conduct of medical imaging?

## METHODS

The key components of our research projects are:

I. **Systematic review:** Collating, evaluating and synthesizing research evidence.
II. **Collecting and analyzing data:** Using retrospectively collected incidents to identify, classify, analyze and review clusters of like-events.
III. **Devising Corrective Strategies:** Prioritizing problems using a combination of methods for searching relevant incidents and using clinical expert review and human factors expertise to devise a set of preventive and corrective strategies.
IV. **Crafting Recommendations:** Crafting materials and a set of recommendations for drawing the attention of healthcare providers to selected problems in the form of alerts.
V. **Pilot Testing:** Testing whether the recommendations are robust, acceptable and feasible by seeking the expert opinions of radiologists and other relevant practitioners; enduring process testing to refine the recommendations.
VI. **Disseminating the refined recommendations:** Disseminating refined recommendations through various dissemination channels, such as emails to individuals, press releases, and information sheets or articles via professional organisations and online dissemination.

## INITIAL RESULTS

So far, our research proposal has been accepted by the UniSA PhD review panel, our ethics has been approved by the UniSA Human Research Ethics Committee (HREC) [Ethics Approval Reference No: 0000034296] issued on 14$^{th}$ Sep, 2015 and we have submitted our systematic review protocol to Joanna Briggs Institute (JBI).

## PLANNED NEXT STEPS

We have started to apply for access to incidents data from different States and Territory health departments. We shall also commence analysing an existing data-source i.e. Radiology Events Register (RaER) of the Royal Australian and New Zealand College of Radiologists (RANZCR). A result will be drawn once we complete our data analysis. We have just finished collecting data through survey and interviews with healthcare professionals, including radiologists, radiographers, and consultants the Annual Scientific Meeting (ASM), 2015 organized by the RANZCR.

## PROPOSED CONTRIBUTION

- ➢ Recommendations and advice for radiologists and others working in medical imaging, thus bridging the gap between HIT and human performance to reduce error and harm in healthcare.
- ➢ Safety recommendations in medical imaging as a means for translating evidence into clinical practice via recommendations for radiologists and other clinicians.
- ➢ To enhance patient safety communication amongst different institutes like University of South Australia (UniSA), the Royal Australian and New Zealand College of Radiologists (RANZCR) and the Australian Patient Safety Foundation (APSF) with respect to strategies to improve patient safety in medical imaging.
- ➢ Capacity building in the disciplines of safety and quality, risk, and human factors for radiologists and clinicians.

# Title of PhD Project: An investigation of adoption of IT outsourcing decision-making research in practice

PhD Student
Mohammad Mehdi Rajaeian
University of Southern Queensland
mohammadmehdi.rajaeian@usq.edu.au

Principal Supervisor
Prof Aileen Cater-Steel
University of Southern Queensland
aileen.cater-steel@usq.edu.au

Associate Supervisor
Dr Michael Lane
University of Southern Queensland
michael.lane@usq.edu.au

## MOTIVATION AND RESEARCH QUESTIONS

While over the last three decades academics have suggested several decision models to assist practitioners with their IT outsourcing (ITO) decision making, there is some evidence that the adoption of those decision models by practitioners has seldom occurred. To date, limited, if any, academic work exists to understand why the research-practice gap in ITO decision making exists and there are few suggestions as to how can the gap can be reduced. This study investigates: RQ1) where and how do practitioners adopt ITO decision-making knowledge? RQ2) what are the barriers of adoption of ITO decision-making research in practice? And RQ3: How can the adoptability of ITO decision-making research by practitioners be increased?

## METHODOLOGY

The research uses a mixed method approach comprising interview-based case studies and online surveys to collect empirical data from ITO practitioners, IT consultants and academic researchers.

## INITIAL RESULTS

Case study in four large Australian organisations confirmed the lack of use of academic research in IT sourcing decisions. The findings revealed that practitioners mainly rely on the advice and lessons learned from 'peer groups', 'IT consultants' and 'Vendors'. The initial findings confirmed the notion of 'two worlds' of research and practice with distinct characteristics as suggested by Caplan's (1979) two-communities theory. However, the propositions of the two-communities theory cannot fully explain the gap. For instance, all the practitioner participants had a university degree and even postgraduate degrees and access to academic research outputs. Therefore difference in the 'language' is not a major hinderance of research use and improving communication between researchers and academics seems a simplistic and not very promising solution. The initial results are aligned with Luhmann's (1982) social system theory, thus the distinction between research and practice is systematic and institutionalised rather than cultural. Interviews with three academic researchers and three IT consultants are also conducted and perceptions of them regarding the factors contributing to the gap between ITO research and practice identified.

## PLANNED NEXT STEPS

In the next phase, two surveys will be developed based on relevant theories and findings from the first phase. The surveys are planned to be administered in March 2016. At the next step quantitative data analysis will be performed and a framework/theory of research-practice gap will be presented.

## PROPOSED CONTRIBUTION

The outputs of this research will be 1) a model/framework that shows the channels and mechanisms of adoption of decision making knowledge by ITO practitioners 2) A framework to explain the research-practice gap in the ITO discipline. The framework contributes to knowledge transfer/utilisation/translation and diffusion theories and aims to identify the factors contributing to the research practice gap in ITO decision making. 3) Recommendations to bridge the gap will be provided. ITO researchers, ITO practitioners and research policy makers will benefit from the outputs of this study. Furthermore, since the research-practice gap is also a problem in other domains of information system, management and various other disciplines, this study can be useful for other researchers studying the research-practice gap in other disciplines.

# Title of PhD Project: Hedonic Information Systems Quality

| PhD Student | Principal Supervisor | Associate Supervisor |
|---|---|---|
| Chanyoung Seo | Zixiu Guo | John D'Ambra |
| UNSW Australia | UNSW Australia | UNSW Australia |
| c.seo@unsw.edu.au | z.guo@unsw.edu.au | j.dambra@unsw.edu.au |

## MOTIVATION AND RESEARCH QUESTIONS

Hedonic information systems (HIS) are an entertainment-oriented information system used in home and leisure settings. It is significantly different from traditional information systems (IS) developed for a user in office environments in which he/she has a specific task to fulfil using the IS. The remarkable success of HIS in a commercial market breaks new grounds for IS research therefore emerging research on HIS emphasises 'fun' and 'enjoyment' felt by its users. Given its practical and academic prominence, this study is captivated to further explore on the topic of HIS, especially HIS quality. With a central focus on enrichment of our current knowledge about HIS, this study is expected to uncover HIS quality attributes from a user perspective. Accordingly, the research questions are formulated as follows;

**$RQ_1$**: What are the attributes of HIS quality from the user perspective?

**$RQ_2$**: What are the antecedents of HIS quality?

## METHODOLOGY

Given that it is in an exploratory nature, this proposal adapts both qualitative - repertory grid technique (RepGrid) for semi-structured interviews - and quantitative approaches - online surveys - in executing the research which will be conducted in three phases. Online games are selected as an study artefact in this study because it is a representative example of HIS that is dominantly used for a recreational purpose and enjoyed by global users.

## INITIAL RESULTS

To date, this research has completed the first round of data collection in Korea, 20 interviews using a RepGrid technique, and content analysis and frequency counts are in progress. The analysis of the collected data utilises a theoretical lens of 'thinking-feeling mechanism' adapted from a balanced thinking-feeling framework. The preliminary findings of this research indicate that HIS quality attributes are compassed by IS functionality-related factors e.g., graphic quality as well as emotional or affective factors e.g., emotion evoked from a user's avatar. Few antecedents of HIS quality e.g., mood, personal interests, and friend group are also identified to be significant in HIS usage contexts. Hence, this study lends support to that HIS quality considerably differs from existing IS quality attributes found in extant IS literature and context-dependency is prominent in IS research.

## PLANNED NEXT STEPS

This research will continue to perform full data analysis of the collected qualitative data. Then, the initial questionnaire online survey will be carried out for the purpose of construct purification, followed by another round of online survey for scale confirmation and model validation.

## PROPOSED CONTRIBUTION

Future findings of this research will have significant contributions to current knowledge on HIS as well as associated industries and practitioners. The present study is one of early efforts in investigating into HIS with a specific focus on quality of fun and pleasurable HIS experience. This research will build both theoretical justifications and empirical support to this area. Hence, it is expected to add insights into a growing body of IS research concerning hedonic sides of IS. In addition, the findings of this study will empower the stakeholders of HIS in an economic sense. With a high level of context-specificity, the findings of this study will provide causal elucidation and cost-effective solutions.

# Title of PhD Project: Data-Driven Services for Predictive Decision-Making in Hospitals

PhD Student
Isabella Eigner
Institute of Information Systems, FAU Nuremberg
Isabella.eigner@fau.de

Principal Supervisor
Prof. Dr. Freimut Bodendorf
Institute of Information Systems, FAU Nuremberg
Freimut.bodendorf@fau.de

## MOTIVATION AND RESEARCH QUESTIONS

As a result of the demographic change and medical advances, demand for healthcare services is increasing steadily. The current supply of resources and qualified employees cannot satisfy this demand (Deloitte 2015; Kick 2005). Using automated, data-driven services to support decision making, hospitals tasks can be performed more efficiently. Although hospitals already have access to a lot of hospital data, the potential of using this information to gather insights for enhanced decision-making is often not fully exploited (Patil et al. 2014). To support decision-making for different stakeholders in hospitals, an integrated system that allows for a valuable and understandable output is needed (Ward et al. 2014). This research aims at developing a clinical decision support system using predictive analytics to aid hospital staff and management in their daily business. Based on initial findings, the main research question is described as follows: *(How) Can the decision making in hospitals be improved by predictive data-driven services provided in an integrated system?*

## METHODOLOGY

As an underpinning framework to answer the proposed question the design science cycles by Hevner (2007) is used. A problem is sought to be solved by designing and developing an artefact, in this case the decision support system and the respective algorithm. In a first step, the environment and important stakeholders in the hospital context are identified (literature review, expert interviews). In a next step, the required data for the following analyses has to be defined, cleaned and structured for further analysis (CRISP-DM Reference model). The prepared data set can then be used to model the proposed services and apply the developed services in an integrated system to support the needs of different stakeholders or applications (Prototyping), e.g., risk management. After deployment and the first evaluation focusing on performance and accuracy, the final artefact is finally submitted to further evaluation considering usability and comprehension in a real life setting through user tests.

## INITIAL RESULTS

Healthcare analytics has been a growing research area for the past few years (Koh and Tan 2005; Raghupathi and Raghupathi 2014), often used for fraud detection (Aral et al. 2012; Christy 1997), and risk prediction (Son et al. 2010) or supporting financial and administrative actions, e.g. by reducing patient length-of-stay (Kudyba and Gregorio 2010). One application of healthcare analytics is clinical decision support, which aims at providing insights to clinical providers by "disseminating timely, actionable information" (Strome 2015) at the right place and time. Therefore, relevant stakeholders can be supported with relevant, intelligently filtered information at appropriate times, to enhance health and health care." (Osheroff et al. 2007) To determine the right suggestion to the relevant stakeholders, predictive analytics methods are used to extract patterns from historical data to create empirical predictions and methods for assessing the quality of those predictions in practice (Shmueli and Koppius 2010).

## PLANNED NEXT STEPS AND PROPOSED CONTRIBUTION

The goal of this research is the design and development of a clinical decision support system based on predictive data-driven services for hospitals. With the support of such a system, pressure on personnel could be reduced, leading to fewer errors and a higher quality in patient treatment as well as lower costs for hospital management. As a next step I aim to finish my fundamental literature review and theoretical grounding until the mid of next year and subsequently start the data preparation. For this purpose I intend to conduct expert interviews at an Australian hospital group to gather first insights on relevant issues within this context.

# Title of PhD Project: Customer and Firms Interaction on fan pages

| PhD Student | Principal Supervisor | Associate Supervisor |
| --- | --- | --- |
| Hamidreza Shahbaznezhad | Arvind Tripathi | Ananth Srinivasan |
| Researcher | Associate Professor | Professor |
| h.shahbaznezhad@auckland.ac.nz | a.tripathi@auckland.ac.nz | a.srinivasan@auckland.ac.nz |

## MOTIVATION AND RESEARCH QUESTIONS

Organisations seeking to improve user engagement continue to invest in fan pages on social media platforms. They expect and plan that in a progressive global marketplace users will utilise these technologies, which will enhance organisational competitive advantages. However, fewer than 1% of fans actively engage with brands online (Nelson-Field et al., 2012). Also we don't know very much about the characteristics of those users whose engage in fan pages. Though a fan page has become an integral part of any firm's social media strategy, we do not yet to understand how firms' fan page strategies affect user engagement. Further, functionality of the social media platforms is often underutilised and is rarely understood as a new enabler for customer and firm interactions. The diversity of users' motivations is considered to be another factor determining user fan page engagement. The variety among organizations' fan page strategies, social media platform characteristics and users' heterogeneity leads to different patterns of user engagement. Correspondingly, the diversity of firms' business, reputation, size and diversification have made managing the fan pages challenging. In this regard, we answer three research question in our study as below.

1. How to characterize firm's user engagement strategies on fan pages?
2. What are the common characteristics of the users which are considered important for the company on fan pages?
3. What drives user participation on social media fan pages for firms and brands?

## METHODOLOGY

The general approach to data analysis would be a deductive data driven approach. Quantitative data can be collected to measure user engagement, such as number of like/favourite, or number of retweet/share of company content in a specific time span. The plan for analysing qualitative data would be content analysis. Opinion mining is the main method for transforming qualitative data to quantitative data. For the first research question, by taking the advantages of ANOVA method, we evaluate heterogeneity among various users in a large data set. We have selected K-Means clustering approach for identifying different types of users that are participating in firm's fan pages. For the second research question we use sentiment analysis and topic modelling for finding the differences between users which are in the centre of concentration of the company and other users. The results of analysis first and second research questions provides us a result that lets me develop an empirical model to explain the users' adoption of diverse strategies. In order to extract and organize tacit knowledge behind user participation, Structural Equation Modelling (SEM) will be the data analysis method for the third research question. By presenting a model, we intend to find the relation between the users' motivations, enablers of social media and firm strategies to analyse user engagement.

## INITIAL RESULTS

We found 6 clusters of user strategies on fan pages in two prominent airline companies in pacific region. We found that firms tend to follow those active users that their general activity and number of followers in platform is higher than others such as opinion leaders.

## PLANNED NEXT STEPS

The last phase of data collection and survey design is under progress.

## PROPOSED CONTRIBUTION

This research aims to contribute to the literature on social media by proposing a comprehensive user engagement framework on firm fan pages. We first identify types of user engagement strategies on fan pages. So the contribution in this step is discovering the online behavioral patterns. Secondly we find the firm sensitivity to the topics and sentiments on fan pages. Finally by uncovering existing behavioral and technological patterns in the commercial-based communication, we systematically analyze those factors that affect user engagement. These results provide an ontology of the underlying components of customers and firm relationship from a socio-technical point of view. As a product of the whole research, online users' learning would be the contribution to the literature.

# Title of PhD Project: Visual representation of the patient pathway to enhance quality of care

PhD Student
Vishakha Sharma
Federation University
vishakhasharma@federation.edu.au

Principal Supervisor
Andrew Stranieri
Federation University
a.stranieri@federation.edu.au

Associate Supervisor
Sally Firmin
Federation University
s.firmin@federation.edu.au

## MOTIVATION AND RESEARCH QUESTIONS

Patients in the health care system with the same condition experience diverse and almost unique sequences of events known as patient pathways or journeys. Although depictions of patient pathways are emerging for health analytics and for visualising electronic health records, the capacity for visual pathway representations to enhance discussion amongst diverse health care professionals in multi-disciplinary meetings (MDMs) has not been explored. Reviews have identified that MDM's can enhance the quality of care, however many meetings are not appreciably deliberative. This research aims to explore the impact a visual representation of a patient's journey has on MDM discussions.

The main research question is: *How can visual representations of patient's pathways facilitate group reasoning in MDMs?*

## METHODOLOGY

The research question will be answered using both quantitative and qualitative analyses. The research will proceed in two phases; a qualitative piloting phase and a quantitative experimental phase. During the first phase, transcripts of discussions typically 4-8 minutes long in three separate MDM's associated with cancer care in Victoria will be recorded and analysed for the Walton and Krabbe discourse types. In addition, the content of the dialogues will be analysed using a thematic analysis technique. Participants from the MDM's will also be interviewed and their responses will be thematically analysed. This is expected to reveal features of conventional MDM discussions sourced from approximately 12 patients. For half of the patients, a visual pathway will be constructed prior to the MDM from health records. For the remaining patients, the presenting physician will only narrate a brief case history.

The results of the first phase will be used to refine the design of an experiment in the second phase, designed to answer a specific hypothesis that may be: *The presentation of visual representation of the patient's pathway will improve deliberation within an MDM.*

A control group comprises multiple MDM meetings discussing colorectal cancer patients. An experimental group comprises MDM meetings discussing cases of similar complexity. Cases in the experimental group will have a visual representation presented to the meeting and displayed throughout the discussion. The control group will not have a visual representation. Patients will be randomly selected from those discussed in MDM's to form the control group. Patients that match control group patients on key criteria will be selected to form the experimental group. Measures of deliberation quality will be used as measure of deliberation activity.

## PLANNED NEXT STEPS

I have a plan for coming three years and expect to complete it by the end of 2018. The coming year will be spent on designing the system. We have already designed a few templates manually and will be finalising one soon. By the end of the next year, I will start evaluating the effectiveness of this visual representation. Last year will be spent on completion of thesis and preparing towards participating in an International Conference.

## PROPOSED CONTRIBUTION

My contribution involves the discovery of the extent to which the representation of a patient's visual pathway can enhance deliberation in multi-disciplinary meetings surrounding cancer care.

# Title of PhD Project: Bridging Learning Analytics with Xorro-Q: An Institutional Dashboard for Engagement

**PhD Student**
Shadi Esnaashari
Massey University
S.Esnaashari@massey.ac.nz

**Principal Supervisor**
Dr Anuradha Mathrani
Massey University
A.S.Mathrani@massey.ac.nz

**Associate Supervisor**
Prof Paul Watters
Massey University
P.A.Watters@massey.ac.nz

## MOTIVATION AND RESEARCH QUESTIONS

Many studies have emphasized the low participation rates in classes and have argued that students need to participate in group activities in class in order to learn more. Different technologies have been used by researchers in classroom environments to increase the participation of students. However, low participation of students seems to be a real issue in many university classrooms. This study will investigate low participation issues using an audience interaction tool named Xorro-Q in the classroom. Xorro-Q enables anonymous interaction between students and the lecturer. This study, therefore, aims to enhance student participation and learning. The motivation behind this study is to see how the engagement of students in classes can increase when they use audience interaction tools and investigate how increasing the engagement of students in the class is related to their course outcome. Two research questions are posed: 1) Does students' participation improve by using audience participation tools such as Xorro-Q? 2) How early can we identify students who are probably close to drop out?

## METHODOLOGY

The methodology used in this study is action research. Action research aims to solve current practical problems by involving active involvement of the study group (community) with underlying theoretical concepts. Using Xorro-Q as an intervention tool for engaging students in conjunction with Activity Theory elements, the study will investigate students' participation in classrooms. Both quantitative and qualitative data will be collected. Quantitative data collected through Xorro-Q will report data about student attendance, participation rate of students in different activities and their scores. Qualitative data will be collected through open-ended survey questions and interviews with lecturers and students.

## INITIAL RESULTS

An exploratory study has been conducted with an audience interaction tool in two undergraduate university classroom settings. In one classroom setting which was a first year computing course, the lecturer employed continuous informal discussion-based teaching activities with Xorro-Q tool. The other classroom setting involved a second year computing course where Xorro-Q was used for formally assessing students' subject knowledge using traditional methods. Our aim was to examine how students engaged with the tool in classrooms. The preliminary findings showed that the audience participation tool has a promising direction for engaging students in the process of learning. Findings indicated that the class which adopted continuous informal discussion approach rendered more enjoyment among students, although the traditional formal assessment activities showed higher student participation.

## PLANNED NEXT STEPS

The next phase of the study will involve further data collection with use of the audience engagement tool. Machine learning algorithms will be used to analyse the data collected to understand how participation of students can be increased in classrooms and if predictions can be made on student grades based on their class participation rates. This will be further supplemented with interview data, to provide social context to the learning analytical methods.

## PROPOSED CONTRIBUTION

There is no evidence in literature examining the relationship between student participation through audience interaction tools and student outcomes. It is hoped that this study will offer new insights on the role of audience interaction tool and has innovative pedagogical implications for teaching and learning practices. Findings of this study will help tertiary institutes to identify students which are close to drop out and find ways to help them.

# Title of PhD Project: Online Community of Practice: Leveraging Information Technology for Reflective Practice in Promoting Mathematical Literacy

PhD Student
Zaenal Abidin
Massey University, Auckland, New Zealand
Z.Abidin@massey.ac.nz

Principal Supervisor
Dr. Anuradha Mathrani
Massey University, Auckland, New Zealand
A.S.Mathrani@massey.ac.nz

Associate Supervisor
Dr. David Parsons
The Mind Lab, Unitec
New Zealand
david@themindlab.com

Associate Supervisor
Dr. Suriadi Suriadi
Massey University, Auckland, New Zealand
S.Suriadi@massey.ac.nz

## MOTIVATION AND RESEARCH QUESTIONS

The advent of information and communication technology (ICT) has brought significant changes in education domain, especially in the field of mathematics education. Teachers are exploring ICT strategies, as they revisit and evaluate how classroom teaching can be aligned to real-world contexts. The national mathematics curriculum of Indonesia mandates that teachers should be able to integrate ICT within instruction and also address external challenges, such as the Programme for International Student Assessment (PISA) results. A 2012 PISA assessment study ranked Indonesian students' performance in the field of mathematics to be second lowest in the league table. This indicates that mathematical literacy skills of Indonesian students are in need of much improvement, although it may be noted that PISA assessments does have a fair share of critics.

An online community of practice (CoP) will be built to foster the process of promoting mathematical literacy. The online CoP supports collaborative learning through exchange of ideas and share classroom experiences to improve teacher's performance. This is a form of reflective practice. The CoP platform has been selected because it has proven to be successful in teacher professional development.

Three research questions are posed: (1) How do teachers engage with ICT tools to design mathematical literacy-type tasks with mathematics curriculum to promote mathematical literacy? (2) How do online CoP facilitate reflective practice amongst teachers for promoting mathematical literacy?, and (3) How can technology-enabled teaching delivery be made more effective to promote mathematical literacy?

## METHODOLOGY

This study will employ action research methods. Five research participants' teaching practices will be investigated in promoting mathematical literacy by engaging with online CoP as a means for reflection. The C4P (Content, Conversation, Connections, Context and Purpose) framework will be used to assist in illustrating the extent to which technology tools provide opportunities for developing social learning. Both qualitative and quantitative data will be gathered. Quantitative data will be collected through questionnaires, while qualitative data will be collected through open-ended survey questions, classroom observation, discussion forum posts (Facebook) and semi-structured interviews.

## INITIAL RESULTS

Findings from the preliminary study conducted in April 2015 revealed that teachers mostly do not know much about mathematical literacy and how technology intervention strategies are applied to mathematics curriculum. This PhD study is thus motivated by the desire to promote mathematical literacy in Indonesia and to increase teachers' appreciation for the significance of mathematical literacy among students with the aid of ICT.

## PLANNED NEXT STEPS

The next steps are designing the mathematical literacy-type tasks through formative evaluation approach, teacher workshop and developing online CoP, and implementing action research to gather both qualitative and quantitative data.

## PROPOSED CONTRIBUTION

This research will add to the body of knowledge theories and practices related to the use of ICT to promote mathematical literacy. Based on study findings, a CoP framework utilizing the 4 C's will be developed. This will help in theorizing concepts related to technology-enabled community of practice-based reflective practice for teachers in Indonesia and globally.

# Title of PhD Project: A Problem Solving-centric Approach to Data Analytics

PhD Student
SHIANG-YEN TAN
QUT
s40.tan@qut.edu.au

Principal Supervisor
Taizan Chan
QUT
t.chan@qut.edu.au

Associate Supervisor
Yue Xu
QUT
yue.xu@qut.edu.au

## MOTIVATION AND RESEARCH QUESTIONS

Most of the data analytics systems to date have been focused on data exploration, a process which helps users to understand their data. However, the complex and interconnected nature of problems faced by practitioners often requires them to go beyond understanding the data technically, to undergo higher level analytical processes such as synthesizing individual analyses, generating and testing hypotheses. Existing systems are lacking of the features that support these high-level analytical processes, and often failed to delivery actionable insights that can be pragmatically applied to solve a real-world problem. This study proposes that data analytics systems which explicitly support high level analytical processes can lead to analytical outcome that is more readily to inform real-world decision making. The primary objective of this study, therefore, is to develop a design theory that can inform the design of integrated data analytics systems that support the entire analytical processes.

## METHODOLOGY

The analytical processes in this study is decomposed into three problem-solving stages, namely, 1) data exploration, 2) information synthesis, and 3) structured reasoning. A conceptual framework is developed to understand users' analytical needs from the perspectives of analytics, informatics, decision, and cognition in each of the three stages. Subsequently, the framework is used to derive design guidelines which provides prescriptive statements to which how the needs can be meet through different functionalities. This study involves a two-stage evaluation. Firstly, a preliminary evaluation involves validating the framework and its components at conceptual level. In the second stage, a working prototype that operationalizes the design guidelines is to be evaluated in an experimentation.

## INITIAL RESULTS

In the preliminary evaluation, interviews with expert panels who are mainly from business and law domains have affirmed that the supports for high-level analytical process are helpful to support and consistent with what has been previously done in the user's mind or by paper-and-pen. The main concern is that these processes themselves are ill-structured and can be very complex. Key findings show that externalization of these mental activities in the systems might entail the possibilities that 1) very complex user inputs and interactions are required, 2) not all kinds of mental reasoning activities can be supported, and 3) the system's framework might constrain the user's flow of thoughts.

## PLANNED NEXT STEPS

A user experimentation is planned to be conducted by first quarter of 2016 to evaluate the working prototype. The experimentation intended to examines the problem solving performances of users, in comparison with the use of conventional data analytics systems. The performance is mainly evaluated in terms of insight quality (i.e. depth and accuracy, comprehensiveness, and practicality of the findings). For the system to be useful, the gains in the insight quality should outweigh the trade-off in terms of the extra time and efforts spent on the analytical task.

## PROPOSED CONTRIBUTION

As the theoretical contribution, the design theory can help researchers 1) to understand user needs and design considerations in the analytical processes as a whole and its individual stages, 2) to understand how the provision of the supports to these needs can influence the user's analytical performance.

Practically, the design theory and the prototype developed can be a guideline for informing the development of systems of similar class. The long term endeavor of this study is to enable the matured version of the system to be freely available to the research community.

# Title of PhD Project: Sharing information in the primary health sector: An examination of the causal mechanisms that govern the inter-organisational appropriation of an ICT-enabled collaborative tool.

PhD Student
Stuart McLoughlin.
Swinburne
smcloughlin@swin.edu.au

Principal Supervisor
Helana Scheepers
Swinburne
hscheepers@swin.edu.au

Associate Supervisor
Rosemary Stockdale
Swinburne
rstockdale@swin.edu.au

## MOTIVATION AND RESEARCH QUESTIONS

This proposed enquiry seeks to explore the appropriation of an Information and Communications Technology (ICT) designed to foster the collaboration logic within the Australian public health sector through the lens of critical realist (CR) based affordance theory. In tandem with an institutional agenda of partnership and collaboration, ICT are seen as pivotal in increasing the efficiency and cost-effectiveness of the Australian primary health sector. New information technologies and increased information access should increase collaborative efficiency and result in better primary healthcare quality. Yet the implementation of collaborative ICT's in this sector lags. This research proposal would ask the question why. The capacity to distribute information and knowledge to collaborative partners (Cho and Mathiassen 2007), with varying practices and differing information priorities, to ensure that any such information and knowledge is assimilated and integrated (Weber and Khademian 2008) is not a straightforward matter. Our knowledge of how actions, contexts and institutional logics impact on ICT adoption patterns, collaboration and information sharing in the Australian primary health sector, is limited.

The purpose of CR is to establish causality. An 'affordance' is the possibility for action arising from a technical artefact (Volkoff and Strong 2013) and, a causal structure triggering organisational change when actualised . This CR-based approach allows us to explore the role of the material in organisational change. To date the CR IS literature has focused on the role of affordances and how they interact with agency to explain organisational outcomes. However, it is argued that affordances, in themselves, cannot provide a complete explanation of how the relationship between agency and structure results in organisational outcomes that arise from the adoption and diffusion of ICT innovations (Bygstad et al. 2015; Volkoff and Strong 2013). The interaction between societal and organisational causal structures and agency needs to be more fully explicated as part of the relationship between organisation and affordances that can trigger change. We need to ascertain how both the organisational and technical structures condition action in this context. This causal-based approach dictates that the researcher must ask what mechanisms – both technical and organisational - must exist in order to account for what can be observed (Easton 2010). Thus the research question is formulated as follows: *What are the mechanisms that explain the collaborative appropriation of a primary health care activity classification information system at the organisational and inter-organisational level?*

## METHODOLOGY

A multiple case-study of events in which users, in a number of public health organisational settings, seek to develop ICT-enabled cross-organisational information sharing, is envisaged. Consistent with critical realism, a retroductive analysis utilising narrative networks is proposed to examine patterns of ICT appropriation. Here it is proposed to examine appropriation patterns that arise from the interaction between organisational and technical causal structures that arise from the implementation of a digital artefact seeking to enable disparate organisations within the public health sector to collaborate through the sharing of information. The artefact is a multidimensional public health activity classification system that seeks to produce sharable digital data in order to promote standardisation and thus disseminate information in a readily understandable form. The artefact holds the potential to improve planning and network coordination and cooperation between collaborating organisations.

## PLANNED NEXT STEPS

The analytical process starts with a thick description of the structures - events and the sequence of events, the actions, powers, liabilities and relationships being observed, that lead to the outcome under investigation. The study framework is to be based on Wynn and Williams (2012) 5 methodological principles for conducting critical realist case study research. Instruments and protocols for interviews and focus groups are to be developed for this pilot project phase.

## PROPOSED CONTRIBUTION

This research will investigate the inter-relationship of non-technology related generative mechanisms with IT-based affordances on ICT usage patterns, and thus seek to contribute to the development and extension of critical realist based affordance theory at the inter-organisational level. This research extends the use of IS theory to the collaborative dynamic of the public health sector. The application of critical realism based affordance theory to the adoption and diffusion of a public health IS innovation will provide practical knowledge on the organisational and inter-organisational impact of ICT in the public health sector. IS authors have argued that affordance theory, and its underpinning philosophy, critical realism, can assist with contextual insights that will help practioners better manage the use of IS innovations. (Mingers et al. 2013; Mutch 2010; Strong et al. 2014; Volkoff and Strong 2013; Wynn Jr and Williams 2012). Given the importance of this sector for civil society and the economy, it is becoming more and more essential to understand how public health actors interact with collaborative ICT's at an inter-organisational level.

# Title of PhD Project: Exploring Formative Assessment in Virtual Learning Environments

PhD Student
Chinthake Wijesooriya
The University of Queensland
c.wijesooriya@business.uq.edu.au

Principal Supervisor
Jon Heales
The University of Queensland
j.heales@business.uq.edu.au

Associate Supervisor
Peter Clutterbuck
The University of Queensland
p.clutterbuck@business.uq.edu.au

## MOTIVATION AND RESEARCH QUESTIONS

Formative assessment has been instrumental in improving learning (Duchesne et al. 2013), but the nature of formative assessment in a contemporary educational environment is yet to be fully investigated. Increased dependency, and the dynamic nature of technology, together with changed learning styles and learning environments, highlight the importance of virtual learning environments (VLEs) in supporting formative assessment. Beyond observation and experience, contemporary learning is characterised by the use emails, social media, and online forums. Often technology enabled communication and informal social groups are significantly effective in learning (Waters and Gasson 2012). The dynamic nature of technology development, extensive use of mobile technology, and use of social media contributes to wide range of learning styles. This research examines the effectiveness of different forms of formative assessment in an undergraduate introductory information systems course. Accordingly, the research question for this research is, "What are the effective forms of formative assessment in a virtual learning environment?"

## METHODOLOGY

To capture the current nature of VLE and current formative assessment practices, we conducted a number of exploratory pilot studies. Data, relevant to student's perceptions of their VLE and the attributes of formative assessment at different stages throughout the learning process was collected. These early results allowed us to formulate our research model and methodology.

Following the identification of different forms of formative assessment attributes, a longitudinal survey method will be used to collect data. Longitudinal data collection provides an opportunity to collect time series data that allows and examination of how formative assessment processes change over time, and the effect of those formative assessment activities.

## INITIAL RESULTS

Preliminary results from pilot studies show that students use email as a preferred feedback information source. Students also use social media for learning throughout the semester.

## PLANNED NEXT STEPS

The research consists of two pilot studies and a longitudinal survey. The two pilot studies will explore virtual learning environment and formative assessment practices in learning environments. These have been completed and data analysis is in progress. Based on the outcome, refining the research model and reviewing the project details will be done during next few months. Longitudinal data will be collected from a tertiary educational environment during next two semesters. The data collection will be completed by the middle of 2016. Any unforeseen issues will be managed by planning an additional round of data collection during the second semester 2016. Writing the thesis will be an ongoing task and the completion of the thesis will be approximately June 2017.

## PROPOSED CONTRIBUTION

This research contributes to theory by explaining and understanding what aspects of formative assessment are effective in contributing to positive learning outcomes. A model for effective formative assessment will be developed and tested. From a practitioner perspective, the outcomes from this research will assist in the effective design and development of formative assessment tasks in VLE's.

# Agency Theory and information sharing in organizations: effects of Locus of Control of Information, and individualism/collectivism (Research-in-Progress)

Photchamarn Prai-ngam (Ph.D.student)
Dr.Stephen Wingreen (Supervisor)
The Department of Accounting and Information Systems
University of Canterbury
Christchurch, New Zealand
Email: ney.prai-ngam@pg.canterbury.ac.nz, stephen.wingreen@canterbury.ac.nz

## Summary

Information sharing in organizations or across multiple agencies remains a critical challenge for CEOs and CIOs worldwide, and the obstacles to information sharing have long been addressed and recognized. Even though differences between information systems, technologies, platforms, and personal information rights pose challenges, there are also hidden problems, which may be related to lack of trust, organizational and individual outcomes, motivation, and goal setting. Agency Theory predicts that the efficiency of decision making is related to the efficiency of information systems in firms ("information systems" is used in the most generic sense, and does not require the use of information technology), especially in its optimal contract form, in which one person, the principal, delegates work to another, the agent. In the context of information sharing, the quality of the decisions to share information is directly related to the efficiency of the organization's information system. A general assumption of Agency Theory is that agents make better decisions when they reap the rewards of their good decisions, and pay a price for their poor decisions.

To the extent that employees with "internal" Locus of Control (LOC) believe that their own decisions affect their rewards and reinforcements in any given context, then in the context of information sharing, this research theorizes that LOC for information sharing will positively affect motivation to share information, and the quality of the decision to share information. The "Individualism – Collectivism" variable represents an individual's cultural inclination toward either self-autonomy on the one hand, or the will of a collective, on the other. To the extent that a self-autonomous employee would make decisions in his own interest, rather than that of the collective, and to the extent that "internal LOC" represents an individual belief in self-control over rewards and reinforcements, then this research theorizes that Individualism – Collectivism will affect the motivation and decision to share information, and the quality of decisions to share information.

This research is expected to make contributions to both theory and practice. First of all, with regard to theoretical contributions, there appears to be little or no research on the topic of how Agency Theory applies to information sharing in governmental organizations. If this research is successful, it will represent a significant expansion of Agency Theory. Since Agency Theory provides a rich theoretical language to describe the behaviour of decision-making agents, direct comparisons between managers and decision-makers in governmental organizations and their counterparts in private firms may be made, analysed, and specified in the language of Agency Theory. This research will also contribute to the literature on Agency Theory by expanding the knowledge about how LOC and individualism-collectivism operate in the domain of decision-making in the context of information exchange and sharing. Finally, this research will expand the theory and measurement of LOC, by developing and introducing a new LOC scale for information exchange.

The practical significance of this research will be realized by providing a systematic means of understanding, coordinating, and organizing the decision structures of governmental organizations. Since individualism-collectivism has a strong cultural component, this research will be of practical significance to governmental agencies across a variety of cultures.

# Title of PhD Project: Towards Discovering the Role of IS Capabilities in Disruptive Innovations

PhD Student
Abayomi Baiyere
TUCS, University of Turku.
abayomi.baiyere@utu.fi

Principal Supervisor
Hannu Salmela
University of Turku, Finland
hannu.salmela@utu.fi

Associate Supervisor
Tomi Dahlberg
Abo Akademi, Turku, Finland
tomi.dahlberg@utu.fi

## MOTIVATION AND RESEARCH QUESTIONS

When Disruptive Innovations occur, they challenge the fabric, structure and capability that define any organization facing their threat. The ability of an organization to restructure and reconfigure its resources to face or leverage such turbulent situations is dependent on how it can orchestrate its resources and capabilities. One essential capability of today's organisations is their IT/IS capability. This research is therefore set up to investigate the different roles an organisation's IT/IS capabilities play in disruptive innovation scenarios. The study explores the significance of IT as an enabler, a sustainer and even as a barrier in a disruptive innovation scenario. Using IT/IS Capability as a theoretical lens, the research questions on which this study will be pivoted are: (a) *In what ways does an organization's IS capabilities influence its actions in disruptive innovation scenarios?* (b) *How can IS capabilities be leveraged to advance an approach to dealing with disruptive innovation?*

## METHODOLOGY

The research method to be adopted for this work will be a qualitative research based on multiple case study approach with the aim to provide an in-depth investigation of the different roles played by IS Capabilities in the occurrence of a disruptive innovation phenomenon. Building on this, the research subsequently adopts a design science approach to advance a framework (MIND Capability Canvas) which is currently being evaluated in the cases as a sense-making device for assessing the IS capability of an organization.

## INITIAL RESULTS

The present results indicate the approaches and IS capability attributes of the case organisations. Correspondingly, areas of similarities have been identified and mapped during the preliminary cross-case analysis carried out on the data. Observations emerging from these interviews reveal striking insights on how each organisation is leveraging their respective IS capabilities differently to attend to the disruptive innovation. For instance, while company AB seems to be actively striving to keep its IT organisation and processes flexible enough for quick adaptation to change (by adopting modular and less restrictive approaches), company CD has adopted a strategy to keep its IT organisation and processes stable/structured and consequently it appears to be rather robust but change averse. It worth noting that during the course of this research, one of the case companies, which a few years back, held a leading and dominant position in its industry, has faltered into a staggering decline due to the introduction of a disruptive innovation in its industry. Another company in an adjacent industry has recently acquired the company. On the other hand, another similarly responding case company has been able to pivot its business and leveraged on its IS capability among other capabilities to keep surviving the disruption. The data collection and analysis has been carried out with the lens of a unified view of IS capability. In order to assess the IS capability of organisations, an IT artefact tagged the MIND Capability Framework has been developed and is currently being evaluated.

## PLANNED NEXT STEPS

Case interviews are planned to continue. In addition, evaluation workshops using the developed IS Capability MIND framework are to be conducted to enable a robust assessment of selected organisation's IS capability.

## PROPOSED CONTRIBUTION

Contributions can include: a) Giving more insight into how IS capability has been leveraged by organizations that have created/responded to disruptive innovations. b) Identifying the difference between the IS capability of proactive or reactive DI organizations. c) Providing a deeper understanding of what the significant role of IS capability has been to organizations facing disruptive innovations. Summarily, the research should generate multidisciplinary theoretical contributions and provide a sense-making framework to inform and guide practitioners.

# Title of PhD Project: Effects of Personalised Web Search on Higher Education Students

| PhD Student | Principal Supervisor | Associate Supervisor |
|---|---|---|
| Sara Salehi | Helen Ashman | Tina Du |
| University of South Australia | University of South Australia | University of South Australia |
| sara.salehi@mymail.unisa.edu.au | Helen.Ashman@unisa.edu.au | Tina.Du@unisa.edu.au |

## MOTIVATION AND RESEARCH QUESTIONS

Students' heavy dependency on general-purpose Web search and its effect on their learning process could create new opportunities to improve learning or cause unexpected pedagogical problems. This phenomenon motivated us to look at personalised Web search as a widespread educational tool. We want to find out:

1. What proportion of students use search engines as their primary or even only source of information? And which search engine they predominantly use?
2. Using the Google search engine, confirmed by research question 1, as the most popular search engine among students, what is the magnitude of difference between personalised and non-personalised search in terms of actual search results for academic queries?
3. In an educational context, what is the role of personalised Web search in improving students' gratification or satisfaction?
4. What parameters/factors in personalised search influence students' outcomes?
5. How do Web search biases influence students' interactions with search engines? Are these bias manifestations different in personalised search as opposed to non-personalised search results?

## METHODOLOGY

### Part I: Survey
A survey was designed to answer research question 1 by investigating information-seeking attitudes of 120 university students and the role that general-purpose search engines play in their education.

### Part II: Experiment
In order to identifying the difference between personalised search and non-personalised search results (research question 2), two sets of 120 informational academic queries were simultaneously submitted to the Google search engine: once anonymously and once logged in with a Google account. The first 10 search results of each search, 240 searches in total, were collected. For each query, the personalised search result list was compared to the non-personalised result list.

### Part III: Study Sessions
During the sessions students will complete a number of controlled and self-directed Web search tasks while the computer screen is being recorded. Then, they will fill out a short questionnaire about their experience after completing each task.
First, we investigate the level of students' satisfaction with both personalised and non-personalised Web search. Second, we observe how much Web search biases influence student's decision making process. Third, how much of a difference personalisation makes in the search results of actual students submitting the exact same query.

## INITIAL RESULTS

Thus far, we learned that the majority of students prefer Google to other search engines; indeed sometimes it is their primary or only academic information-seeking tool. Additionally, about 80% of them use search engines for educational purposes on daily basis. Our results also showed that on average only 53% of links appear, not necessarily in the same order, in both personalized and non-personalized search results. Interestingly, we observed only slight differences in the extent of personalization based on academic topics.

## PLANNED NEXT STEPS

The main part of this work, Part III of the methodology section, is still at the data collection stage and data analysis and interpretation of this part is the next step and the focus point of the doctoral consortium discussions.

## PROPOSED CONTRIBUTION

As distance and traditional education are becoming more self-driven, the answers to this research questions make a noteworthy contribution to search engine studies and are important to education providers.

# Title of PhD Project: Exploring Online Engagement: How Online Communities Utilise Multiple Social Networking Platforms

PhD Student
Rhys McIlwaine
Victoria University of Wellington
rhys.mcilwaine@vuw.ac.nz

Principal Supervisor
Dr. Jocelyn Cranefield
Victoria University of Wellington
jocelyn.cranefield@vuw.ac.nz

Associate Supervisor
Prof. Pak Yoong
Victoria University of Wellington
pak.yoong@vuw.ac.nz

## MOTIVATION AND RESEARCH QUESTIONS

Creating serial video content is becoming an increasingly lucrative business opportunity for content creators, who are using online communities to sustain their multi-million dollar advertising partnerships. To engage in these online communities members utilise multiple social networking platforms. In these platforms, there are specific social and technical norms which impact engagement across the whole community. Currently there is little research into how these single platform norms enable and constrain engagement across online entertainment communities.

In response to this opportunity, there are two research questions in this study. The first is to understand how individuals utilise the range of social networking platforms when engaging in online communities. These individuals include both the content creators and other members. The second research question seeks to understand the impact that the perceived affordances and constraints of social networking platforms have on engagement. This question will be analysed both in terms of social and technical affordances.

## METHODOLOGY

Aligned to an interpretive worldview at an individual level of analysis, this study will utilise the netnographic process (a sub-set of virtual ethnography) to answer the research questions. This includes the collection of three different types of data: an observational journal from the perspective of the researcher to support research question one; interviews which seek to expand on the ideas in the observational journal and support research question two; and social media data which is the primary source of data for this study and supports both research questions.

Data will be collected over two phases. In the first phase, social media data for three months of engagement by the content creator will be collected manually and by a social media aggregator. An observation journal based on the content creator's social media will be produced in this phase. Within the second phase, three months of engagement data from other community participants will be collected by a social media aggregator, and this will inform the second element of the observation journals. Interview data will also be collected at this time.

Based on these three data sources, structuration theory will be utilised to interpret the data to understand how the social and technical norms of multiple social networking platforms affect engagement in online communities.

Future steps for this research include completing the first data collection phase and refining the design for the second research phase being conducted next year.

## PROPOSED CONTRIBUTION

This research is intended to contribute specifically to theory development, information systems methods, and to practice. The most important contribution of this research to develop theory with respect to multiple platform engagement by individuals within online communities from structuration perspective. This theory development will explain how recursion between social and technical norms on multiple platforms impacts engagement in online entertainment communities.

Additionally, this research contributes to information systems methods by highlighting the use of netnography; a method which is underutilised in information systems research given the centrality of the IT artefact to its process. This research also applies structuration theory to a new context, helping to highlight future research opportunities

Finally, this study intends to contribute to the practice by informing content creators both how individuals and the platforms that are used inform online community engagement. This can be used to help content creators manage their social media platform portfolio and their personal engagement practices.

# Title of PhD Project: A Social Influence Theory of Participation in the Service Co-creation through Social Networks.

PhD Student
Reihaneh Bidar
QUT
reihaneh.bidar@hdr.qut.edu.au

Principal Supervisor
Dr. Jason Watson
QUT
ja.watson@qut.edu.au

Associate Supervisor
Prof. Alistair Barros
QUT
alistair.barros@qut.edu.au

## MOTIVATION AND RESEARCH QUESTIONS

The primary aim of this research is to understand the potential engagement of members in service co-creation through online social networks. The research focuses on two main areas; the potential social factors in customer to customer (C2C) relationships among members, and the structure of the social network that is critical to the formation of communities and spread of behaviour through the network. Therefore, the following research questions have been developed:

RQ1: How is participation in service co-creation influenced by social factors?
RQ2: How is participation in service co-creation influenced by the network structure itself?
RQ3: What are implications of social factors and network structure on participation in service co-creation through social networks?

## METHODOLOGY

This research will use a multiple-holistic case study. The desired cases must be co-creation driven social networks that provide and deliver services from the user-base. The data will be collected by conducting interviews, observations and social network analysis (SNA) to address research questions. Firstly, a semi-structured interview will be conducted to know the views, experiences and ideas of participants so that social influences within the context of co-creation can be explained in depth. To increase the reliability and accuracy of subjects' responses, the research will use observation by tracking members' previous and live activities (e.g. shared information). Thematic analysis will be used to analyse the interview and observation data and connect overlapping codes by using NVivo. Using inductive analysis approach, the results will answer how social factors influence participation in the co-creation of services through social networks. Secondly, snowball sampling and mapping through observation and interview will be conducted to obtain a visual representation of the network. Using the representative graph of the network, the researcher will combine and understand the network narratives to describe how network structures are linked to participation in co-creation. The process will be done by understanding the structure of subjects. The outcome of SNA will answer how the network structure influences participation by understanding the connectivity between members and structure of their relationships. Finally, the result of the interview, observation and SNA will provide a model which represents implications of social factors and network structure on participation in service co-creation through social networks.

## PLANNED NEXT STEPS

The next steps for this research will be the refinement of a conceptual model until December 2015, starting data collection in December 2015 for six months following the analysis of the data.

## PROPOSED CONTRIBUTION

This study contributes to understanding the role of social influence and its effects on user participation of service co-creation through social networks. This study will make valuable contributions to the area of co-creation in the social network, both theoretically and practically. From the theoretical perspective, the findings of this research contribute to the foundation of future developments and improvement of social influence theory and co-creation theory, by having a particular focus on service-oriented social networks. The study will conceptualise user collaboration on service co-creation based on the affected social factors. From the practical perspective, this research will give insight into attracting more co-creators by practitioners and enhance their business model. Furthermore, the research might inform the design of future architectures of participation in different utility platforms.

# Title of PhD Project: Managing Risks in ERP Projects

| PhD Student | Principal Supervisor | Secondary Supervisor | Associate Supervisor |
| --- | --- | --- | --- |
| Mark Van der klei | Tyron Love | Trevor Nesbit | Lloyd Carpenter |
| University of Canterbury | University of Canterbury | University of Canterbury | Lincoln Canterbury |
| Mark.vanderklei@canterbury.ac.nz | Tyron.love@canterbury.ac.nz | Trevor.nesbit@canterbury.ac.nz | Lloyd.Carpenter@lincoln.ac.nz |

## MOTIVATION AND RESEARCH QUESTIONS

Enterprise Resource Planning (ERP) solutions offer organisations significant benefits regarding resource planning, resource allocation, and strategic reporting functionality. However, before these benefits can be realised these systems must first be installed. Although systems that integrate multiple business units (such as ERPs) have been in existence for over 3 decades, an inability to manage ERP risks during installation remains at the forefront of ERP project management issues. Part of this can be attributed to the complex interconnected nature of ERP risk factors, where risks occurring early in an implementation have the potential to influence different risks later in that same implementation. In addition, contrary findings about how risks can be controlled (which includes singular and portfolios of control) have contributed to the formative state of theory-based research examining the relationship between risks and controls at the project implementation level.

Therefore the following research questions arise:

How does the relationship between different risks change during the different stages of an ERP implementation?

How can Project Managers map risks to controls across different stages of ERP implementations?

## METHODOLOGY

Due to the difficulty found with identifying the boundaries between ERPs and their implementation contexts, it is vital that ERP projects be examined in real live settings. In such challenging situations, any method incapable of acknowledging the complexities of the environment, the inter-connectedness and subsequent inter-relationships between risks and their associated controls in ERP implementations would be unsuitable. As an indigenous researcher, I have chosen to utilise an indigenous methodology (Kaupapa Māori) which incorporates into its core a series of key principles which include the inter-connected nature of people and their environments, and ethical considerations regarding respect and reciprocity. In addition to fulfilling some of the more challenging aspects of this research, these core principles reflect my ontological beliefs and axiological values. To enhance the methodological fit, Action Design Research (ADR) has been chosen as the method for gathering data within a Kaupapa Maori framework.

## INITIAL RESULTS

A promising finding is that of a two-way hierarchical structure. This structure consists of different risks and controls hierarchical connected to each other, but in different directions. Unresolved risk factors can result in the manifestation of other risks later in the project, and it was found that controls for specific risks can be mapped in the opposite direction to risks, resulting in the formulation of portfolios of control.

## PLANNED NEXT STEPS

Testing the initial findings in a live ERP project are the planned next steps, in conjunction with interviews with experienced ERP practitioners.

## PROPOSED CONTRIBUTION

The researcher seeks to expand on what is known in relation to the management and control of risks in ERP projects, and to develop a framework capable of informing and directing both experienced and new to ERP project participants on risk management activities in these complex projects.

# Title of PhD Project: Sensemaking Processes and Stakeholders' Reactions to IT Implementation

PhD Student
Amir Rahighi
Swinburne University
arahighi@swin.edu.au

Principal Supervisor
Professor Judy McKay
Swinburne University
jmckay@swin.edu.au

Associate Supervisor
Dr Rajiv Vashist
Swinburne University
rvashist@swin.edu.au

## MOTIVATION AND RESEARCH QUESTIONS

Among the identified reasons and causes that can undermine IT adoption and implementation success, resistance to change, negative attitudes and dysfunctional behaviours, are recognised as salient impediments to efficient and effective change implementations that need to be understood and addressed. Understanding stakeholders' reactions to change and knowing why and how resistance to IT implementation occurs and develops, gives change agents the opportunity to direct the implementation process and change in a way that is more successful and productive. In contrast, if resistance to change is not considered and addressed appropriately, negative feelings and interpretations develop and dysfunctional and destructive behaviours and reactions occur, and ultimately it may undermine the IT implementation endeavours and end in project failure. In this regard, an objective of this study is to derive insights into stakeholders' reactions to change and improve and extend the current understanding of resistant attitudes and behaviours during IT implementations.

The purpose of the research is to investigate, elaborate and understand how the sensemaking (how action and interpretation affect each other) of change agents (CAs) and change recipients (CRs) influences and shapes their reactions (attitudes and behaviours) to change during IT implementations. This research pursues to derive insights into and depict a holistic view of these interpretations and actions among stakeholders, their context, and their consequences in order to understand and interpret developed patterns of responses and reactions, including resistance to change and support. The research questions to assist in achieving this objective are: (RQ1) How do CAs and CRs make sense of IT implementations over time? (RQ2) What are the implications of the sensemaking of CAs and CRs on their reactions to IT implementations?

## METHODOLOGY

This research adopts the interpretive paradigm and follows social constructionism as its epistemology to investigate and understand stakeholders' reactions through a sensemaking lens. The research design of this study is the interpretive multi-case study that is analysed through qualitative methods. This research investigated two case studies and used different sources of evidence (including semi-structured interviews and diaries) for data collection, which allows comparison and contrast between the cases. The researcher studied the process of IT implementation and investigated sensemaking activities between CAs and CRs at three points in time (prior to, during and after the IT implementation). This strategy allows the researcher to study social phenomena and investigate complex interpretive processes over a period of time in order to capture their dynamics and explore changing conditions. This research uses Grounded Theory data analysis method and adopts the coding and categorisation techniques to analyse collected data.

## PLANNED NEXT STEPS

At the current stage of this study, the data collection rounds are completed and all the collected data are prepared for data analysis. Data analysis started after conducting the first round of data collection, and the initial coding of the first case study is finished. Data analysis will be my main focus during next three to five months. Writing and completing the thesis is also a significant activity of the researcher up to the end of this PhD project.

## PROPOSED CONTRIBUTION

Theoretically, this study contributes to the understanding of sensemaking and its implication on CAs and CRs attitudes and behaviours during IT implementations. The research also contributes to practice by developing the understanding of stakeholders' reactions to change. It intends to provide practical suggestions to managers and change agents for managing resistance to change and promoting stakeholders' engagement and support.

# Title of PhD Project: Barriers and Incentives for Cybersecurity Intelligence Sharing Between Public and Private Sector

| PhD Student | Principal Supervisor | Associate Supervisor |
|---|---|---|
| Farzan Kolini | Dr. Lech Janczewski | Dr. Fernando Beltran |
| University of Auckland | University of Auckland | University of Auckland |
| f.kolini@auckland.ac.nz | l.janczewski@auckland.ac.nz | f.beltran@auckland.ac.nz |

## MOTIVATION AND RESEARCH QUESTIONS

The subject of cybersecurity has been widely discussed by authorities, organizations, media, academia and individuals over the past few years. The complexity of cybersecurity operations has made it ultimately impossible for any entity, organization or the government, to protect the cyber assets appropriately without leveraging collaboration with other parties. Therefore, a sustained cybersecurity intelligence sharing (CIS) can reinforce the understanding of the cyber threat landscape and assist the sharing parties with a timely attack detection, response, and risk mitigation capability. Since CIS is just beginning to appear in organisations recently, little research and few insights exist to guide a successful development and implementation of such systems. To fill this gap, this study aims to explore the challenges and incentives that may promote or impede cybersecurity intelligence sharing initiatives between public and private sectors. The number of organisations that are using CIS for cybersecurity purposes is growing, hence determining the determinants that are associated with CIS activities are concerning. Therefore, this research attempts to provide answers for the following research questions: (1). How cybersecurity intelligence sharing can be used to defend against cyber attacks? (2). How public and private sectors describe the barriers and incentives for cybersecurity Intelligence sharing? (3). What might explain differences in cybersecurity intelligence sharing between public and private sectors? (4). How to improve the engagement of public and private sector in cybersecurity intelligence sharing program?

## METHODOLOGY

This study will be using a Sequential Mixed Method approach for designing this study. This strategy will conduct a qualitative study at the beginning for the exploratory purpose and following up with a quantitative method with a large sample that can help to generalize results to the population. The Data collection technique for this study will occur in two distinct steps. The first step proceeds with qualitative data collection technique, open-ended interviews, and it will be followed by the second step, which is quantitative data collection. This study will adapt Monfelt's 14-layerd framework as the theoretical lens to explore the interaction between social and technical aspect of cybersecurity intelligence sharing.

## INITIAL RESULTS

|  | Environment-Social | Organizational | Technical |
|---|---|---|---|
| Incentives | <ul><li>Institutional pressure</li><li>Political support</li><li>Competitive advantage</li></ul> | <ul><li>Increased productivity</li><li>Lower administration burden</li><li>Increased decision making</li></ul> | <ul><li>Quality of information</li><li>Usefulness of information</li><li>Saving time</li></ul> |
| Challenges | <ul><li>Environment complexity</li><li>Legal issues</li><li>Administration & coordination</li><li>Privacy and confidentiality</li><li>Lack of trust</li></ul> | <ul><li>Resistance to change</li><li>Diversity</li><li>Lack of process</li><li>Management attitude</li><li>Goal alignment</li></ul> | <ul><li>Lack of resource</li><li>Data incompatibility</li><li>Confidentiality</li></ul> |

## PROPOSED CONTRIBUTION

The first significance of the study is to provide the insights and wisdom required for the practitioner in this field. Secondly, for theoretical significance, this study attempts to complement the body of knowledge on cybersecurity in general and cyber intelligence sharing by using Monfelt's 14-layerd framework. A mixed-method research proposed will also be of significance in a quantitative -dominated Information Security research in Information System field. It will produce a triangulation of results between both quantitative and qualitative method employed, and provide a richer understanding on the topic being investigated.

# Title of PhD Project: Study to Investigate Factors Influencing Adoption of Mobile Devices in the Health Care Environment

PhD Student
Vasundhara Rani Sood
USQ
Vasundhara.Rani@usq.edu.au

Principal Supervisor
Prof. Raj Gururajan
USQ
Raj.Gururajan@usq.edu.au

Associate Supervisor
Dr. Abdul Hafeez Baig
USQ
abdualhb@usq.edu.au

## MOTIVATION AND RESEARCH QUESTIONS

Use of mobile devices in healthcare has enhanced the scope of health care services and in the near future will make health services more flexible. However, for adopting mobile devices healthcare professionals (HCPs) feel uncomfortable and adoption of mobile devices in the healthcare environment is slow (Slaper & Conkol 2014; Wu, Li & Fu 2011; Milward, Day, Wadsworth, Strang &Lynskey 2015). This raised a question in researcher mind: What are the factors which make healthcare professional reluctant to use mobile devices in healthcare? Therefore, this research is an attempt to find out the factors influencing adoption of mobile devices from health care professional's perspective. The scope of the study is limited to healthcare professionals especially, doctors and nurses who use mobile devices in the healthcare context.

This research on adoption of mobile devices at the individual level in the healthcare domain is focused on the following research questions:

RQ1: What factors constitute a conceptual framework for adoption of mobile devices in healthcare environment?

RQ2: How self-efficacy influence individual readiness for adoption of mobile devices in healthcare?

RQ3: How relative advantages influence individual readiness for adoption of mobile devices in healthcare?

RQ4: What is the role of age, gender, and experience as mediating factors on the determinants of adoption of mobile devices in healthcare?

## METHODOLOGY

This research will use mix methodology: both qualitative and quantitative approaches sequentially to achieve research objectives. In qualitative research design focus group technique will be used. Focus groups can provide valuable information to determine constructs and to develop hypothesises because in focus groups the responses come after a relaxed, comfortable and enjoyable discussion. On the other hand, quantitative research aims at testing of hypotheses which can be achieved if researcher is able to collect large number of responses. In quantitative research design online survey technique will be used to collect data in this research. To increase the response rate of online survey respondents will be contacted through email and personally prior to being sent the survey questionnaire and follow up emails to request them to fill the survey questionnaire. Thus, in this research a qualitative approach will be used to determine constructs and develop hypothesises and formulate a conceptual framework and quantitative approach will be used to test hypothesises.

## INITIAL RESULTS

This research hypothesizes individual readiness, complexity and social influences are mediated by age, gender and experience for individual intention to adopt mobile devices in healthcare. Furthermore, functional features of mobile devices and compatibility with healthcare processes will be mediated by age and gender.

## PROPOSED CONTRIBUTION

The conceptual model developed in this research will be the first of its type for adoption of mobile devices in healthcare. This model will also serve as a pathway for the healthcare and information technology domain to design information communication tools for healthcare which can improve healthcare practices.

# Title of PhD Project: An Intelligent Forecasting System for Forest Fires in Central Kalimantan, Indonesia

PhD Student
Ariesta Lestari
Faculty of IT Monash University
ariesta.lestari2@monash.edu

Principal Supervisor
Dr. Grace Rumantir
Faculty of IT Monash University
Grace.Rumantir@monash.edu

Associate Supervisor
Prof. Nigel Tapper
School of Geography and Environmental Science, Monash University
Nigel.Tapper@monash.edu

## MOTIVATION AND RESEARCH QUESTIONS

Forest fires have become an increasingly serious environmental problem in Indonesia. These massive fires are threatening flora and fauna diversities, and the health and livelihood of local people. Central Kalimantan is one area in Indonesia that has experienced severe forest fires from early 1980 until now. The threat of future fires is a continuing issue because of a lack of mechanisms to prevent and minimise the occurrence of fires. Until now, fire authorities in Central Kalimantan have only relied on daily hotspot monitoring to monitor the occurrence of fires.

A range of methods have been employed in forest fire prediction systems. These include traditional statistical hypothesis testing, linear regression, classification and regression trees and other methods from machine learning. None of the studies using these methods have taken into account the characteristics of forest fires in Central Kalimantan. To address the research gap, the research questions of this study are formulated as follows:
1. In what way can the key factors in the formation of forest fire influence the escalation of hotspots into forest fire?
2. To what extent can data mining techniques be used to intelligently predict forest fires in Central Kalimantan?

## METHODOLOGY

The research development and design for this study will adapt the design science research methodology (DSRM) proposed by Peffers et al. (2007). DSRM is an iterative process oriented methodology. Each activity in this methodology contains recommendations concerning what to do and the objectives that have to be achieved. It starts by identifying the problem, then defining the objectives of a solution, designing and developing a model, demonstrating and evaluating the model, if needed there are iteration process back to design and development before finally communicating the result.

The design and development for this study will adapt the conceptual process of model building that was introduced in (Pyle 1999). The model development starts from data selection, which consists of analysing the state of the data required for solving the problem. After the appropriate data are selected, the next step is data pre-processing to make sure the data are good enough to build the model. The third step is model development and the last step is model deployment, which involves the validation of knowledge and the examining of knowledge discovered by domain experts.

## INITIAL RESULTS AND PLANNED NEXT STEPS

A comprehensive literature review and personal communication with some experts working with forest fires in Central Kalimantan have been conducted to identify the possible factors contributing to forest fires in Central Kalimantan. These identified factors are used to guide data collection. After the data has been collected and pre-processed the next stage is model development. It starts with finding the best data mining techniques for solving the problem of forest fires from literature review. The task of choosing the data mining algorithm includes selecting methods to be used for searching the pattern in the data such as deciding which models may be appropriate to solve problems in the forest fires domain. The test design specifies that the dataset should be separated into training and testing tests. In the building model activities, the modelling tool is run on the prepared dataset to create one or more models. The model then will be assessed to ensure it meets the data mining criteria and passes the desired test criteria.

## PROPOSED CONTRIBUTION

This research project takes an interdisciplinary approach as it integrates knowledge on fires and data mining. This study proposes a model that can conceptualise the factors contributing to forest fires and give information about future fires. This information will help authorities to minimise the threat of loss and damage caused by fire and to optimise firefighting resources and development projects.

# Title of PhD Project: Understanding the Use of Mobile Technology in a Multidisciplinary Care Team

PhD Student
Pamela Spink
Monash University
Pamela.Spink@monash.edu

Principal Supervisor
Frada Burstein
Monash University
Frada.Burstein@monash.edu

Associate Supervisor
Julie Fisher
Monash University
Julie.Fisher@monash.edu

## MOTIVATION AND RESEARCH QUESTIONS

Traditional health care information systems have limitations and recent developments in the area of mobile technologies have led to renewed interest on the adoption of mobile technology based IT systems in the healthcare sector. Mobile technologies are being increasingly used in the health care sector to deliver high quality care to patients through improved communication and contributing to the efficiency and effectiveness of doctors, nurses and allied health professionals. Research into mobile technology in a healthcare setting (hospital) has been fragmented. Contributions to this research area largely focus on benefits of a particular mobile device; adoption/acceptance and attitudes/intentions of use by a particular healthcare provider as well as the barriers to adoption of this technology. There is very limited research undertaken to understand how mobile technology, as defined in this research to include PDAs, mobile phones/smartphones and tablet PCs (but not limited to), can contribute towards supporting communication and interaction in a multidisciplinary care team in a health setting. The proposed research study seeks to understand the use of mobile technology in a multidisciplinary healthcare team and its contribution to improve health outcomes. The following broad research question is being addressed: "How does mobile technology support a multidisciplinary care team in health settings?" and more specifically,

**RQ1.** What are the tasks that are performed by various healthcare professionals in a multidisciplinary care team?
**RQ2.** What are the interactions that take place between the different healthcare team members?
**RQ3.** How and what type of mobile technology is used by different healthcare team members in different context?
**RQ4.** What type of tasks/activities are supported using these mobile technologies in health settings?

## METHODOLOGY

Theoretically, a combination of concepts from activity theory and task technology fit is being used as a lens to generate a more specific model to explore the use of mobile technologies in a multidisciplinary care team in a health setting. Using a qualitative approach, data is being collected through interviews, observations and documents at the case study organization, a large hospital in Melbourne.

## INITIAL RESULTS

Preliminary findings reveal that mobile technology is being used for communication, data management and accessing information.

## PLANNED NEXT STEPS

- Complete data collection by end of February 2016 and commence interpretation, refine model and conclude
- Commence thesis writing by October 2016 and start working on a journal paper.
- Thesis submission and journal paper publishing (2017)

## PROPOSED CONTRIBUTION

For theory, an extended and more specific set of factors based on the chosen theoretical lens that can help explain the use of mobile technology in a multidisciplinary care team. In practice the findings of this study will have important implications for healthcare providers, policy makers and health service researchers.

# Title of PhD Project: Personally Controlled Electronic Health Records (PCEHR) Adoption and Utilization Behavior in Australia

Amirhossein Eslami Andargoli
Swinburne University of technology
aeslamiandargoli@swin.edu.au

Diana Rajendran
Helana Scheepers
Swinburne university of Technology
drajendran@swin.edu.au
hscheepers@swin.edu.au

Amrik Sohal
Monash University
Amrik.Sohal@monash.edu

## MOTIVATION AND RESEARCH QUESTIONS

Health care faces many challenges, with an escalation in demand for more health care services, leading to an on-going increase in the costs associated with their services. Developments through Information and Communication Technology (ICT) provide an opportunity for the health care industry to improve its performance. Electronic health (e-health) is a term that was first used in the 1990s to explain any use of computerized technology in the health care domain. The ultimate goal of e-health is to improve the quality and efficiency of health care services. E-health promises many benefits for the health care industry; however, without a reasonable rate of e-health technology adoption, potential benefits will not be realized. The personally controlled electronic health record (PCEHR) is a cornerstone of the e-health system in Australia; it is an online summary allowing health care providers and hospitals to view and share an individual's health information, including diagnoses, allergies and medications (Royle et al. 2014). More importantly, user engagement and adoption plays a pivotal role in the success of the PCEHR system in Australia. The following research questions will be addressed in this study:

RQ1: How does different stakeholders' behaviour contribute to enable the adoption of PCEHR? Or resist the adoption of PCEHR? What are their implications?

RQ2: What are the critical factors that impact the adoption of PCEHR?

RQ3: How can stakeholder theory explain PCEHR adoption process?

## METHODOLOGY

The case study is the most popular qualitative method used in IS (Myers et al. 1997). This method is suitable for studying a phenomenon in contexts and understanding its causal mechanism (Easton 2000) as cited in Cho 2007). The case study method is suitable for understanding the context of IS and the process over time of mutual influence between the system and its context (Walsham 1993). Therefore, the case study method fits well with the scope of this research to study the contemporary phenomenon of the PCEHR in its context.

## PLANNED NEXT STEPS

This research comprises four main phases. In the first phase the critical factors in adoption of e-health technologies are identified based on a comprehensive literature review (this phase has been completed). The outcome of this phase provides the overview of critical factors in adoption of e-health technology in a global context. In the second phase, adoption of PCEHR in Australia as a single embedded (multiple unit of analysis) case study will be studied through interviewing different stakeholders. The data from this phase will be used in phase 3 for the development of a conceptual framework in the Australian context. In phase 4 the findings of this research will be presented to a focus group in order to obtain feedback and triangulate the results and validate the framework.

## PROPOSED CONTRIBUTION

Health care is a fundamental part of society and in this day and age challenges in relation to the adoption of e-health technologies are very common all around the globe. Overall, research in this domain can contribute to a broader understanding of the critical factors that determine their adoption or otherwise, to process the concerns about adoption amongst stakeholders through appropriate dialogue and thus mitigate impediments in e-health adoption. This PhD will provide a deeper understanding of the issues of adoption of the PCEHR in Australia.